\documentclass[aps,prd,groupedaddress,showpacs]{revtex4-1}
\usepackage{amssymb}
\usepackage{amsmath}
\usepackage{graphicx}

\begin{document}

\date{\today }
\title{On the oscillation-driven cosmological expansion at the post-inflation stage}
\author{Vladimir A. Koutvitsky}
\email{vak@izmiran.ru}
\author{Eugene M. Maslov}
\email{zheka@izmiran.ru}
\affiliation{Pushkov Institute of Terrestrial Magnetism, Ionosphere and Radio Wave
Propagation (IZMIRAN) of the Russian Academy of Sciences,\\ Troitsk, Moscow
Region, 142190, Russia}
\date{\today}

\begin{abstract}
Dynamics of the inflaton scalar field oscillating around a minimum of the
singular potentials in the expanding Universe is investigated. Asymptotic
formulas are obtained describing the cosmological expansion at the late
times. The problem of stability of the oscillations considered and the
related phenomenon of the field fragmentation are briefly discussed.
\end{abstract}

\maketitle

\section{Introduction}

According to the standard inflationary scenario the accelerated expansion of
the Universe occurs when the inflaton scalar field $\phi $ is in the
slow-roll regime on a sufficiently flat part of the potential $V(\phi )$ 
{\cite{Star1, Linde1, Albrecht, Linde2}}. 
{Such potentials can arise when taking into account quantum corrections 
in the right-hand side of the Einstein equations \cite{Star1}}.
As the minimum of the potential is
approached, the speed of the rolling-down increases, and the scalar field
eventually enters the stage of rapid oscillations \cite{Kofman}. For
the potentials having a power-law behavior at the minimum these
oscillations have been considered by Turner \cite{Tur}, who has derived the
effective equation of state for them. Later on, Damour and Mukhanov \cite%
{Dam} have pointed out that inflation will continue for a time at the
oscillation stage too if the potential is non-convex in regions not too far
from the minimum. They also have found a nice geometric interpretation of
this fact and estimated the amount of inflation occurring at the oscillation
stage. This estimation was being improved successively by Liddle and
Mazumdar \cite{Liddle} and Sami \cite{Sami}.

At the present paper the primary emphasis is on the potentials admitting
oscillation-driven inflation, but having singularities at their minima.
Singular potentials are intensively discussed in the modern cosmology in the
context of the so-called sudden singularities occurring in high derivatives
of the metric when the scalar field $\phi $ passes through the minimum of $%
V(\phi )$ (see, e.g., \cite{Barrow1} and references therein). In contrast,
we consider the effect of singularities of the inflaton potentials on the
cosmological expansion dynamics averaged over the oscillation time scale.
The paper is organized as follows. In Section 2, based on the method of
averaging, we give a rigorous treatment of the scalar field oscillations in
the Friedmann-Robertson-Walker Universe. Using these results, in Section 3
we perform the comparison analysis of the cosmological expansion dynamics at
the post-inflation stage for three generic potentials, two of them being
singular. In Section 4 some remarks are made concerning stability of the
oscillations considered.

\section{Scalar field oscillations in the expanding Universe}

The homogeneous inflaton scalar field $\phi $ in the flat
Friedmann-Robertson-Walker Universe is described by the equation 
\begin{equation}
	\phi _{tt}+3H\phi _{t}+V^{\prime }(\phi )=0,  \label{eq1}
\end{equation}%
where $H=a_{t}/a$ is the Hubble parameter, $a(t)$ is the scale factor. If
one assumes the Universe is filled with a scalar field $\phi $ only, the
evolution of the scale factor $a(t)$ will be governed by the Friedmann
equations%
\begin{equation}
	a_{tt}/a=-
\left( 
	4\pi G/3
\right) 
\left( 
	\rho +3p
\right) 
	,  \label{eq2}
\end{equation}%
\begin{equation}
\left( 
	a_{t}/a
\right) 
	^{2}=
\left( 
	8\pi G/3
\right) 
	\rho   \label{eq3}
\end{equation}%
with the effective pressure and energy density%
\begin{equation}
	p=\phi _{t}^{2}/2-V(\phi ),\text{\quad }\rho =\phi _{t}^{2}/2+V(\phi ).
\label{eq4}
\end{equation}%
From Eqs. (\ref{eq2}) and (\ref{eq3}) it follows that%
\begin{equation}
	\rho _{t}=-3(a_{t}/a)(p+\rho ).  \label{eq5}
\end{equation}

Once the slow-roll stage ends, the inflaton field $\phi (t)$ begins to
execute fast damped oscillations around the minimum of the potential $V(\phi
)$. Equation (\ref{eq1}) describing these oscillations can be treated, in
view of Eqs. (\ref{eq3}) and (\ref{eq4}), as a dissipative dynamical system
written in terms of the independent variables $\phi $ and $\phi _{t}$. Let
us go from these variables to the other ones, $\theta $ and $\rho $,
following the technique of separation of fast and slow motions (see, e.g., 
\cite{Mois}). We set%
\begin{eqnarray}
	\phi  &=&\varphi (\theta ,\rho ),  \label{eq6} \\
	\phi _{t} &=&\omega (\rho )\,\varphi _{\theta }(\theta ,\rho ),  \label{eq7}
\end{eqnarray}%
where $\varphi (\theta ,\rho )$ is a $2\pi $-periodic solution of the
equation%
\begin{equation}
	\omega ^{2}(\rho )\,\varphi _{\theta \theta }+V^{\prime }(\varphi )=0.
	\label{eq8}
\end{equation}%
From Eqs. (\ref{eq4}), (\ref{eq6}), and (\ref{eq7}) it follows that the
first integral of this equation is just the energy density $\rho $,%
\begin{equation}
	\frac{1}{2}\,\omega ^{2}(\rho )\,\varphi _{\theta }^{2}+V(\varphi )=\rho ,
\label{eq9}
\end{equation}%
so that%
\begin{equation}
	\omega ^{-1}(\rho )=\frac{1}{\pi \sqrt{2}}\int_{\varphi _{\min }(\rho
	)}^{\varphi _{\max }(\rho )}\frac{d\varphi }{\sqrt{\rho -V(\varphi )}},
	\label{eq10}
\end{equation}%
where $V(\varphi _{\min ,\text{ }\max })=\rho $. Equations (\ref{eq6})-(\ref%
{eq10}) fully determine the transformation $\left( \phi ,\phi _{t}\right)
\rightarrow \left( \theta ,\rho \right) $.

In order for the above procedure to be self-consistent, Eqs. (\ref{eq6}) and
(\ref{eq7}) must be supplemented by the compatibility condition%
\begin{equation}
	\varphi _{\theta }\theta _{t}+\varphi _{\rho }\rho _{t}=\omega \varphi_{\theta }.  \label{eq11}
\end{equation}%

Equations determining\ evolution of the variables $\rho $, $\theta $ are
derived from Eqs. (\ref{eq5}) and (\ref{eq11}) with the use of Eqs. (\ref%
{eq3}), (\ref{eq4}), and (\ref{eq7}):%
\begin{eqnarray}
	\rho _{t} &=&-2\sqrt{6\pi G}\rho ^{1/2}\omega ^{2}(\rho )\varphi _{\theta}^{2},  \label{eq12} \\
	\theta _{t} &=&\omega (\rho )+2\sqrt{6\pi G}\rho ^{1/2}\omega ^{2}(\rho)\varphi _{\rho }\varphi _{\theta }.  \label{eq13}
\end{eqnarray}%
It should be pointed out that these equations are exact and fully equivalent
to Eq. (\ref{eq1}). The similar, but more cumbersome equations can be
obtained for the polar coordinates of the variables $\phi $, $\phi _{t}$ 
\cite{Ren}.

Integrating $p+\rho =\omega ^{2}(\rho )\varphi _{\theta }^{2}$ over $\theta $
from $0$ to $2\pi $ and denoting $\bar{p}=\left( 2\pi \right)
^{-1}\int_{0}^{2\pi }p(\theta ,\rho )\,d\theta $ we immediately obtain the
Turner's formula for the adiabatic index:%
\begin{equation}
	\gamma (\rho )=1+\frac{\bar{p}(\rho )}{\rho }=\frac{\sqrt{2}\omega (\rho )}{%
	\pi \rho }\int_{\varphi _{\min }(\rho )}^{\varphi _{\max }(\rho )}\sqrt{\rho
	-V(\varphi )}\,d\varphi .  \label{eq14}
\end{equation}%
This formula can be also derived with the help of the action-angle variables 
\cite{Masso}. As seen from Eq. (\ref{eq2}), the adiabatic index determines
dynamics of the cosmological expansion: the Universe will expand with
acceleration if $\gamma <2/3$, and with deceleration if $\gamma >2/3$.

The system of equations (\ref{eq12}), (\ref{eq13}) belongs to the class of
systems with a rotating phase. If the field oscillations are sufficiently
fast in comparison with the rate of the cosmological expansion, i.e., $%
H/\omega \sim \varepsilon \ll 1$, the generalized averaging method will be
applicable for simplification of the system. In general, the method consists
in going from $\theta ,\rho $ to the new, "averaged", variables $\bar{\theta}%
,\bar{\rho}$ with the help of asymptotic power series in $\varepsilon $ \cite%
{Mois,Nayf}. In the lowest order we have $\rho =\bar{\rho}+O(\varepsilon
),\;\theta =\bar{\theta}+O(\varepsilon )$, and evolution equations for $\bar{%
\rho}$ and $\bar{\theta}$ are derived for once by averaging over the period $%
2\pi $ of the right-hand sides of Eqs. (\ref{eq12}) and (\ref{eq13}), which
corresponds to the Van der Pol approximation. As a result we find%
\begin{eqnarray}
	\rho _{t} &=&-2\sqrt{6\pi G}\,\rho ^{3/2}\gamma (\rho ),  \label{eq15} \\
	\theta _{t} &=&\omega (\rho ),  \label{eq16}
\end{eqnarray}%
where we have dropped the bar signs and neglected of the term arising from
averaging of the second term in Eq. (\ref{eq13}). The latter is caused by\
the neglect of the next order term ($\sim \varepsilon ^{2}$) in Eq. (\ref%
{eq15}) resulting in the error in $\rho $ of the order of $\varepsilon ^{2}t$%
. This situation is typical when considering the systems with a rotating
phase in the lowest order approximation (see, e.g., \cite{Mois}). Our
following analysis is based on system (\ref{eq15}), (\ref{eq16}).

\section{The potentials}

\begin{figure}[htb]
\includegraphics[width=0.4\textwidth]{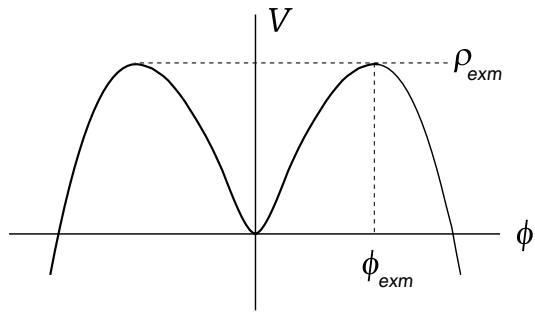}
\caption{Sketch of the shape of potentials (\ref{eq17})-(\ref{eq19}). }
\label{Fig-1}
\end{figure}

Here we will deal with three generic potentials having very similar shapes
(sketched in Fig. \ref{Fig-1}), but essentially different behavior at the minima.
Namely, we examine the canonical $\phi ^{2}-\phi ^{4}$ potential,%
\begin{equation}
	V(\phi )=\frac{m^{2}}{2}\phi ^{2}-\frac{\lambda }{4}\phi ^{4},  \label{eq17}
\end{equation}%
the logarithmic potential,%
\begin{equation}
	V(\phi )=\frac{m^{2}}{2}\phi ^{2}
\left( 
	1-\ln \frac{\phi ^{2}}{\sigma ^{2}}%
\right) ,  
	\label{eq18}
\end{equation}%
and the fractional-power potential,%
\begin{equation}
	V(\phi )=-\frac{m^{2}}{2}\phi ^{2}+\frac{3\lambda }{4}\phi ^{4/3}.	\label{eq19}
\end{equation}%

While the first potential is evidently regular, the two potentials last
named have singularities at the minimum, $\left\vert V^{\prime \prime }(\phi
)\right\vert \rightarrow \infty \;(\phi \rightarrow 0)$. The logarithmic
potentials currently appear in inflationary cosmology \cite{Barrow2} and in
some supersymmetric extensions of the Standard model (e.g., flat direction
potentials in the gravity mediated supersymmetric breaking scenario \cite%
{Enq}). The logarithmic terms arise due to quantum corrections to the bare
inflaton mass. The fractional-power potentials are also considered in
cosmology \cite{Barrow1, Hari}.

Let us investigate the dynamics of cosmological expansion at the stage of
scalar field oscillations for each of the above potentials.

\subsection{The $\mathbf{\phi ^{2}-\phi ^{4}}$ potential}

\begin{figure}[htb]
	\includegraphics[width=0.4\textwidth]{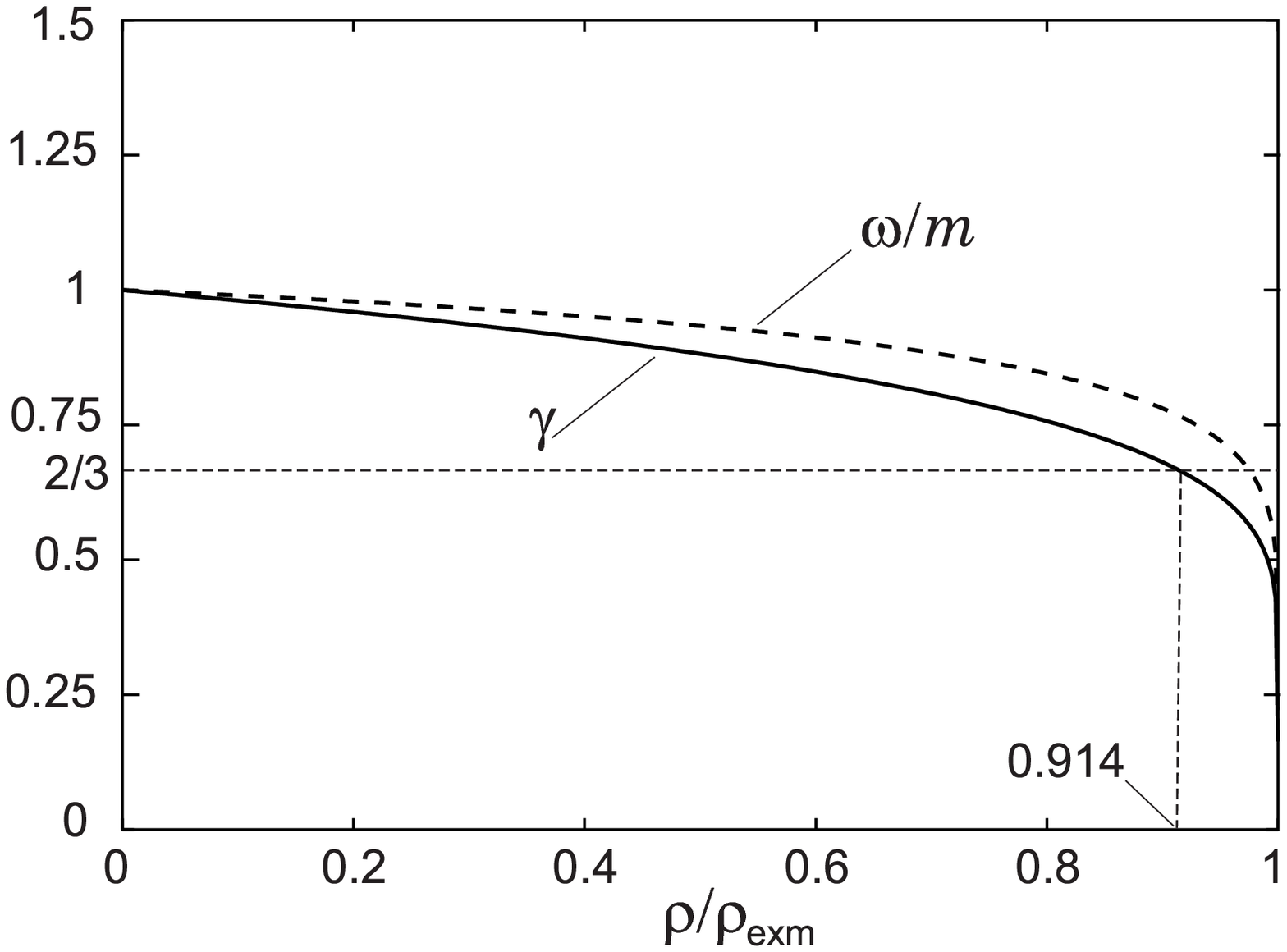}
	\caption{The oscillation frequency and adiabatic index for potential (\ref{eq17}). }
	\label{Fig-2}
\end{figure}

For this potential all quantities can be evaluated exactly. From Eqs. (\ref%
{eq10}) and (\ref{eq14}) we find%
\begin{equation}
	\frac{\omega }{m}=\frac{\pi }{2K(\varkappa )\sqrt{1+\varkappa ^{2}}}, \label{eq20}
\end{equation}%
\begin{equation}
	\gamma =\frac{2}{3\varkappa ^{2}K(\varkappa )}
\left[ 
\left( 
	1+\varkappa^{2}
\right) 
	E(\varkappa )-
\left( 
	1-\varkappa ^{2}
\right) 
	K(\varkappa )
\right]
,  \label{eq21}
\end{equation}%
where $K(\varkappa )$ and $E(\varkappa )$ are complete elliptic integrals, 
\begin{equation}
	\varkappa =\frac{1}{\sqrt{\rho /\rho _{\rm{exm}}}}
\left( 
	1-\sqrt{1-\rho/\rho _{\rm{exm}}}
\right) 
	,  \label{eq22}
\end{equation}%
$\rho _{\rm{exm}}=V(\phi _{\rm{exm}})=m^{4}/\left( 4\lambda \right)
,\;\phi _{\rm{exm}}\equiv \varphi _{\rm{exm}}=m/\sqrt{\lambda }$. Since $%
0<\rho /\rho _{\rm{exm}}<1$, one has $0<\varkappa <1$. Figure \ref{Fig-2}
illustrates the corresponding curves. It is seen that $\gamma =2/3$ is
achieved at the transition point $\rho _{\rm{tr}}/\rho _{\rm{exm}%
}\approx 0.914$.

The damped field oscillations are given by%
\begin{equation}
	\phi (t)=\varphi (\theta ,\rho )=\varphi _{\max }(\rho )\,{\rm sn}
\left( 
	\frac{ 2K(\varkappa )}{\pi }\theta ,\varkappa 
\right) 
,  \label{eq23}
\end{equation}%
where%
\begin{equation}
	\varphi _{\max }=\varphi _{\rm{exm}}\sqrt{\frac{2\varkappa ^{2}}{1+\varkappa ^{2}}}.  \label{eq24}
\end{equation}

At the late times $\tau =\sqrt{6\pi G\rho _{\rm{exm}}}\;t\gg 1$ the energy
density is small, $\varrho =\rho /\rho _{\rm{exm}}\ll 1$, so that%
\begin{equation}
	\frac{\omega }{m}\approx 1-\frac{3}{16}\varrho ,\qquad \gamma \approx 1-%
	\frac{3}{32}\varrho ,  \label{eq25}
\end{equation}%
\begin{equation}
	\frac{\varphi _{\max }}{\varphi _{\rm{exm}}}\approx \frac{\sqrt{2}}{2}%
	\varrho ^{1/2},\qquad \bar{p}\approx -\frac{3m^{4}}{128\lambda }\varrho ^{2}.
\label{eq26}
\end{equation}%
In this case the general integral of Eq. (\ref{eq15}) has the form%
\begin{equation}
	\varrho ^{-1/2}+\frac{\sqrt{6}}{16}\ln \frac{1-\frac{\sqrt{6}}{8}\varrho
	^{1/2}}{1+\frac{\sqrt{6}}{8}\varrho ^{1/2}}+const=\tau .  \label{eq27}
\end{equation}

The constant in (\ref{eq27}) is determined by the initial conditions. It is
seen that their effect is much more essential at the late times than the
impact of the logarithmic term that tends to zero as $\varrho \rightarrow 0$%
. Ignoring the initial conditions and, hence, the logarithmic term, we
obtain asymptotically%
\begin{equation}
	\varrho \sim \tau ^{-2},\quad a\propto \varrho ^{-1/3}\propto \tau ^{2/3}.
\label{eq28}
\end{equation}%
Thus, the scalar field, oscillating around a quadratic-power minimum of the
regular potential, behaves as a non-relativistic matter 
{\cite{Star2}}.

\subsection{The logarithmic potential}

\begin{figure}[htb]
	\includegraphics[width=0.4\textwidth]{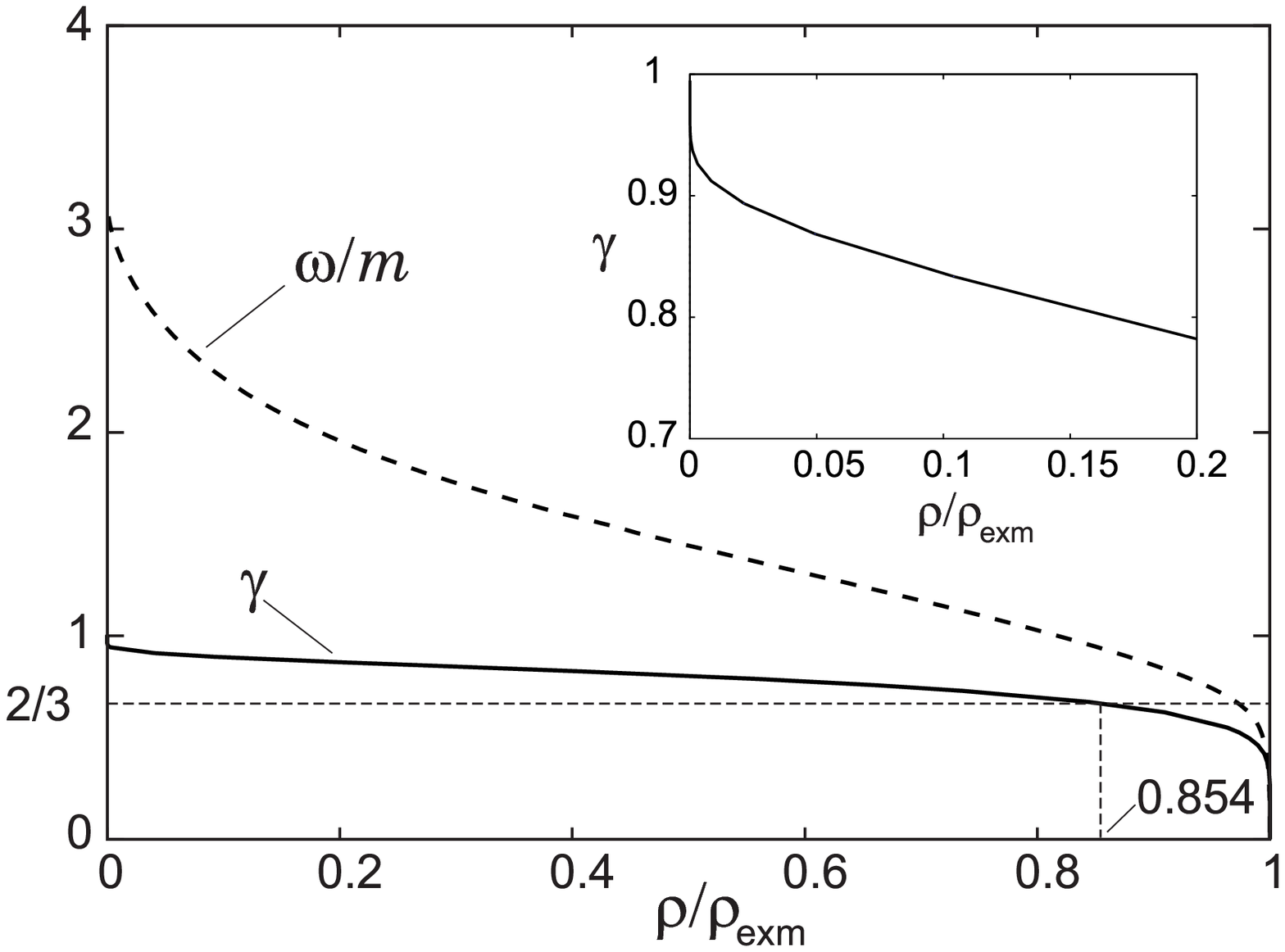}
	\caption{The oscillation frequency and adiabatic index for potential (\ref{eq18}). }
	\label{Fig-3}
\end{figure}

Consider now potential (\ref{eq18}) having weak logarithmic singularity at
the origin. In this case $\rho _{\rm{exm}}=m^{2}\sigma ^{2}/2$, $\varphi _{%
\rm{exm}}=\sigma $, $\rho _{\rm{tr}}/\rho _{\rm{exm}}\approx
0.854$ (see Fig. \ref{Fig-3}). For the late times, $\tau \gg 1$, $\varrho =\rho /\rho
_{\rm{exm}}\ll 1$, from (\ref{eq10}), (\ref{eq14}) we obtain the
asymptotic expansions%
\begin{equation}
	\omega /m=\Delta 
\left( 
	1-a_{2}\Delta ^{-2}-a_{4}\Delta ^{-4}+O(\Delta^{-6})
\right) 
	,  \label{eq29}
\end{equation}%
\begin{equation}
	\gamma =1-b_{2}\Delta ^{-2}-b_{4}\Delta ^{-4}+O(\Delta ^{-6}),  \label{eq30}
\end{equation}%
where $\Delta ^{2}=1-\ln \left( \varphi _{\max }/\varphi _{\rm{exm}%
}\right) ^{2}\gg 1,$ $\Delta ^{2}e^{1-\Delta ^{2}}=\varrho \ll 1$.

\noindent Calculation gives: $a_{2}=1-\ln 2\approx 0.307,$ $a_{4}=\pi ^{2}/8-\ln
4+(1/2)\ln ^{2}2\approx 0.088,$ $b_{2}=1/2,$ $b_{4}=3/4-\ln 2\approx 0.057$.
Restricting ourselves to the leading terms of these expansions we find%
\begin{equation}
	\frac{\omega }{m}\approx \sqrt{-\ln \varrho },\qquad \gamma \approx 1+\frac{1%
	}{2\ln \varrho },  \label{eq31}
\end{equation}%
\begin{equation}
	\frac{\varphi _{\max }}{\varphi _{\rm{exm}}}\approx \sqrt{-\frac{\varrho }{%
	\ln \varrho }},\qquad \bar{p}\approx \frac{m^{2}\sigma ^{2}}{4}\frac{\varrho 
	}{\ln \varrho }.  \label{eq32}
\end{equation}%
In this approximation the general integral of Eq. (\ref{eq15}) is given by%
\begin{equation}
	\varrho ^{-1/2}
\left( 
	1-\frac{1}{2\ln \varrho }+O
\left( 
	\ln ^{-2}\varrho
\right) 
\right) 
	+const=\tau  \label{eq33}
\end{equation}%
Ignoring the effect of initial conditions we obtain asymptotically%
\begin{equation}
	\varrho \sim \tau ^{-2}
\left( 
	1+1/\ln \tau ^{2}
\right) 
	,\quad a\propto \tau ^{2/3}
\left( 
	\ln \tau 
\right) 
	^{1/6}.  \label{eq34}
\end{equation}%
Surprisingly, the field oscillations are still sinusoidal,%
\begin{equation}
	\phi (t)=\varphi (\theta ,\rho )\sim \varphi _{\max }(\rho )\sin \theta ,
	\label{eq35}
\end{equation}%
but with the frequency increasing as $\sqrt{\ln \tau }$. We thus conclude
that expansion dynamics with potential (\ref{eq18}) is very similar to that
with potential (\ref{eq17}).

\subsection{The fractional-power potential}

\begin{figure}[htb]
	\includegraphics[width=0.4\textwidth]{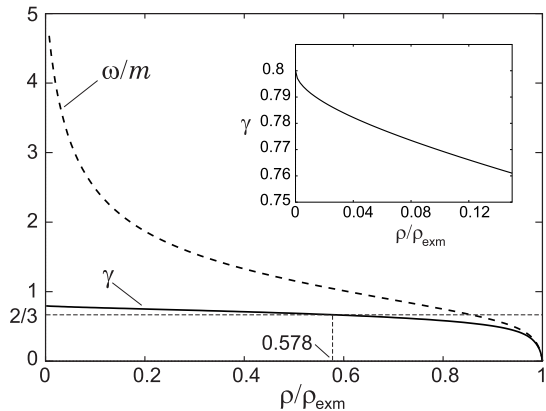}
	\caption{The oscillation frequency and adiabatic index for potential (\ref{eq19}). }
	\label{Fig-4}
\end{figure}

Let us examine potential (\ref{eq19}) having strong fractional-power
singularity. In this case $\rho _{\rm{exm}}=\lambda ^{3}/\left(
4m^{4}\right) $, $\varphi _{\rm{exm}}=\lambda ^{3/2}/m^{3}$, $\rho _{\rm{%
tr}}/\rho _{\rm{exm}}\approx 0.578$ (see Fig. \ref{Fig-4}). 

\noindent The corresponding asymptotic expansions for the late times are given by%
\begin{equation}
	\omega /m=c_{0}\delta ^{-1/3}
\left( 
	1-c_{2}\delta ^{2/3}-c_{4}\delta ^{4/3}+O(\delta ^{2})
\right) 
	,  \label{eq36}
\end{equation}%
\begin{equation}
	\gamma =d_{0}
\left( 
	1-d_{2}\delta ^{2/3}-d_{4}\delta ^{4/3}+O(\delta ^{2})
\right) 
	,  \label{eq37}
\end{equation}%
where $\delta =\varphi _{\max }/\varphi _{\rm{exm}}\ll 1,$ $3\delta
^{4/3}-2\delta ^{2}=\rho /\rho _{\rm{exm}}=\varrho \ll 1,$ $c_{0}\approx
1.071,$ $c_{2}\approx 0.441,$ $c_{4}\approx 0.101,$ $d_{0}=4/5,$ $%
d_{2}\approx 0.174,$ $d_{4}\approx 0.122$. 
{Notice that the above values of $d_{0}$ and $d_{2}$ can be also derived from formula (16) of  Ref. {\cite{Tur}}, 
obtained for the case of a more general potential.} 

Taking into account in (\ref{eq36})-(\ref{eq37}) the leading
terms only, we find%
\begin{equation}
	\frac{\omega }{m}\approx c_{0}
\left( 
	\frac{\varrho }{3}
\right) 
	^{-1/4},\quad \gamma \approx \frac{4}{5}
\left[ 
	1-d_{2}
\left( 
	\frac{\varrho }{3}
\right)
	^{1/2}
\right] 
	,  \label{eq38}
\end{equation}%
\begin{equation}
\frac{\varphi _{\max }}{\varphi _{\rm{exm}}}\approx 
\left( 
	\frac{\varrho }{3}
\right) 
	^{3/4},\;\bar{p}\approx -\frac{\lambda ^{3}}{20m^{4}}\varrho 
\left[
	1+4d_{2}
\left( 
	\frac{\varrho }{3}
\right) 
	^{1/2}
\right] 
	.  \label{eq39}
\end{equation}%
With (\ref{eq38}), the general integral of Eq. (\ref{eq15}) takes the form%
\begin{equation}
	\varrho ^{-1/2}
\left( 
	1-\frac{d_{2}}{\sqrt{3}}\varrho ^{1/2}\ln \varrho^{1/2}+O(\varrho )
\right) 
	+const=\frac{4}{5}\tau ,  \label{eq40}
\end{equation}%
so that we obtain asymptotically%
\begin{equation}
	\varrho \sim \frac{25}{16\tau ^{2}}
\left( 
	1+\frac{5d_{2}}{2\sqrt{3}}\frac{\ln \tau }{\tau }
\right) 
	,\quad a\propto \tau ^{5/6}.  \label{eq41}
\end{equation}%
The field oscillations are essentially non-sinusoidal,%
\begin{equation}
	\phi (t)=\varphi (\theta ,\rho )\sim \varphi _{\max }(\rho )\cos ^{3}\psi(\theta ,\rho ),  \label{eq42}
\end{equation}%
where $\psi (\theta ,\rho )$ is determined by the elliptic integrals,%
\begin{eqnarray}
	&&2E
\left( 
	\psi ,\sqrt{2}/2
\right) 
	-F
\left( 
	\psi ,\sqrt{2}/2
\right)  
	\notag
\\
	&=&
\left( 
	\sqrt{3}/3
\right) 
\left( 
	\omega /m\right) ^{-1}
\left( 
	\varphi_{\max }/\varphi _{\rm{exm}}
\right) 
	^{-1/3}\theta .  \label{eq43}
\end{eqnarray}%
It is seen that in this case the scalar field dynamics and the rate of the
oscillation-driven cosmological expansion differ drastically from those for
potentials (\ref{eq17}), (\ref{eq18}).

\section{Concluding remarks}

The above results were obtained under the assumption that the spatially
uniform oscillations conserve their coherence over a long period of time.
This means these oscillations were assumed to be stable or at least
quasistable. To examine the stability one must proceed, instead of (\ref{eq1}%
), from the full Klein-Gordon equation%
\begin{equation}
	\phi _{tt}+3H\phi _{t}-a^{-2}\Delta \phi +V^{\prime }(\phi )=0.  \label{eq44}
\end{equation}

Consider small perturbations around spatially uniform oscillations,%
\begin{equation}
	\phi (t,\mathbf{r})=\phi (t)+\delta \phi (t,\mathbf{r}),\qquad 
\left\vert
	\delta \phi 
\right\vert 
	\ll 
\left\vert 
	\phi 
\right\vert .  
	\label{eq45}
\end{equation}%
Setting%
\begin{equation}
	\phi (t)=\varphi (\theta ,\rho ),  \label{eq46}
\end{equation}%
\begin{equation}
	\delta \phi (t,\mathbf{r})=a^{-3/2}(\rho )\,\omega ^{-1/2}(\rho )\int Y(\theta,\mathbf{k})e^{i\mathbf{kr}}d\mathbf{k,}  \label{eq47}
\end{equation}%
in the linear approximation we arrive at the equation%
\begin{equation}
	d^{2}Y/d\theta ^{2}+\omega ^{-2}
\left[ 
	k_{\rm ph}^{2}+V^{\prime \prime }(\varphi(\theta ,\rho ))
\right] 
	Y=0,  \label{eq48}
\end{equation}%
where ${k_{\rm ph}}=k/a$ is the physical wavenumber, and the terms of the second order in $\varepsilon \sim
H/\omega $ were neglected. The dependence $a(\rho )$ and slow evolution of $%
\rho (\theta )$ are determined by the equations%
\begin{equation}
	a_{\rho }/a=-
\left( 
	3\rho \gamma (\rho )
\right) 
	^{-1},\quad \rho _{\theta}/\rho =-3
\left( 
	H/\omega 
\right) 
\gamma (\rho ),  \label{eq49}
\end{equation}%
resulting from (\ref{eq3}), (\ref{eq15}), and (\ref{eq16}).

Since $\varphi (\theta ,\rho )$ is a $2\pi $-periodic function in $\theta $
satisfying Eq. (\ref{eq8}), equation (\ref{eq48}) belongs to the class of
the Hill's equations with a slowly varying parameter. If $V^{\prime \prime
}(\varphi )$ becomes infinite periodically, Eq. (\ref{eq48}) will be
singular Hill's equation. For potentials (\ref{eq18}), (\ref{eq19}) the
singular Hill's equations with constant parameters were investigated in
Refs. \cite{Kutv1, Kutv2, Kutv3} where the structure of resonance zones over
the $k_{\rm ph}^{2},\varphi _{\max }^{2}$-plane was revealed. It turns out that for
the sufficiently large $\left( k_{\rm ph}/m\right) ^{2}$ there exists a sequence of
the narrow resonance zones with very small values of the Floquet exponent,
while in the domain $\left( k_{\rm ph}/m\right) ^{2}\sim 1$ one has a wide resonance
zone where the exponent is significantly higher. As the Universe expands, $%
k_{\rm ph}^{2}$ and $\varphi _{\max }^{2}$ decreases. The corresponding points on the 
$k_{\rm ph}^2,\varphi _{\max }^{2}$-plane move along the trajectories representing
evolution of the perturbation $k$-modes for a given initial $\varphi _{\max
}^{2}$. From the start these points cross rapidly the narrow resonance zones
not giving $k$-modes the chance to be significantly amplified. With time the
speed of the points decreases, and eventually the points enter slowly the
domain $\left( k_{\rm ph}/m\right) ^{2}\sim 1$ where the $k$-modes begin to grow
exponentially. At the nonlinear stage this results in decay of the spatially
uniform field $\phi (t)$ into oscillating localized lumps, the pulsons
(oscillons). This phenomenon was observed in numerical experiments in Refs. 
\cite{Enq2, Amin}. Notice that in the case of the logarithmic potential the
scale $\left( k_{\rm ph}/m\right) ^{2}\sim 1$ corresponds to the size of the
gaussian-like pulson being a solution of Eq. (\ref{eq44}) with $H=0$. It is
clear that sufficiently massive pulsons should be considered as
selfgravitating objects, gravipulsons \cite{Kutv4}, disturbing significantly
the Friedmann-Robertson-Walker metric. Thus the results of the previous
sections are valid only up to the beginning of the field fragmentation stage.


\bigskip

\begin{thebibliography}{99}
\bibitem{Star1} 
{A.A. Starobinsky, "A new type of isotropic cosmological 
models without singularity", Phys. Lett. B \textbf{91}, 99 (1980)}.

\bibitem{Linde1} A.D. Linde, "A new inflationary universe scenario: A
possible solution of the horizon, flatness, homogeneity, isotropy and
primordial monopole problems", Phys. Lett. B \textbf{108}, 389 (1982).

\bibitem{Albrecht} A. Albrecht and P. Steinhardt, "Cosmology for Grand
Unified Theories with radiatively induced symmetry breaking", Phys. Rev.
Lett. \textbf{48}, 1220 (1982).

\bibitem{Linde2} A.D. Linde, "Chaotic inflation", Phys. Lett. B \textbf{129}%
, 177 (1983).

\bibitem{Kofman} L. Kofman, A. Linde, and A.A. Starobinsky, "Reheating after
inflation", Phys. Rev. Lett. \textbf{73}, 3195 (1994).

\bibitem{Tur} M.S. Turner, "Coherent scalar-field oscillations in an
expanding universe", Phys. Rev. D \textbf{28}, 1243 (1983).

\bibitem{Dam} T. Damour and V.F. Mukhanov, "Inflation without slow roll",
Phys. Rev. Lett. \textbf{80}, 3440 (1998).

\bibitem{Liddle} A.R. Liddle and A. Mazumdar, "Inflation during oscillations
of the inflaton", Phys. Rev. D \textbf{58}, 083508 (1998).

\bibitem{Sami} M. Sami, "Inflation with oscillations", Gravitation \&
Cosmology, \textbf{8}, 309 (2002).

\bibitem{Barrow1} J.D. Barrow and A.A.H. Graham, "Singular inflation", Phys.
Rev. D \textbf{91}, 083513 (2015).

\bibitem{Mois} N.N. Moiseev, \textit{Asymptotical Methods of Nonlinear
Mechanics }(Nauka, Moskva, 1981), in Russian.

\bibitem{Ren} A.D. Rendall, "Late-time oscillatory behaviour for
self-gravitating scalar fields", Class. Quantum Grav. \textbf{24}, 667
(2007).

\bibitem{Masso} E. Masso, F. Rota, and G. Zsembinszki, "Scalar field
oscillations contributing to dark energy", Phys. Rev. D \textbf{72}, 084007
(2005).

\bibitem{Nayf} A.H. Nayfeh, \textit{Perturbation Methods }(John Wiley \&
Sons, 1973).

\bibitem{Barrow2} J.D. Barrow and P. Parsons, "Inflationary models with
logarithmic potentials", Phys. Rev. D \textbf{52}, 5576 (1995).

\bibitem{Enq} K. Enqvist and J. McDonald, "Q-balls and baryogenesis in the
MSSM", Phys. Lett. B \textbf{425}, 309 (1998).

\bibitem{Hari} K. Harigaya, M. Ibe, K. Schmitz, and T.T. Yanagida,
"Dynamical fractional chaotic inflation", Phys. Rev. D \textbf{90}, 123524
(2014).

\bibitem{Star2}
{ A.A. Starobinsky, "On the nonsingular isotropic cosmological model", 
Sov. Astron. Lett. \textbf{4}, 82 (1978).}

\bibitem{Kutv1} V.A. Koutvitsky and E.M. Maslov, "Parametric instability of
the real scalar pulsons", Phys. Lett. A \textbf{336}, 31 (2005).

\bibitem{Kutv2} V.A. Koutvitsky and E.M. Maslov, "Instability of coherent
states of a real scalar field", J. Math. Phys. (N.Y.) \textbf{47}, 022302
(2006).

\bibitem{Kutv3} V.A. Koutvitsky and E.M. Maslov, "The singular Hill equation
and generalized Lindemann-Stieltjes method", J. Math. Sci. \textbf{208}, 222
(2015).

\bibitem{Enq2} K. Enqvist, S. Kasuya, and A. Mazumdar, "Inflatonic solitons
in running mass inflaton", Phys. Rev. D \textbf{66}, 043505 (2002).

\bibitem{Amin} M.A. Amin, R. Easther, H. Finkel, R. Flauger, and M.P.
Hertzberg, "Oscillons after inflation", Phys. Rev. Lett. \textbf{108},
241302 (2012).

\bibitem{Kutv4} V.A. Koutvitsky and E.M. Maslov, "Gravipulsons", Phys. Rev. D \textbf{83}, 124028 (2011).
\end{thebibliography}
\end{document}